\documentstyle[12pt,aps]{revtex}
\begin{document}
\begin{center}
{\Large \bf  On the Entropy of Minimally Coupled and Non-minimally Coupled 
     Gravities}
\end{center}
\vspace{1cm}
\centerline{{Jeongwon Ho\footnote{E-mail address : jwho@physics.sogang.ac.kr},
Yongduk Kim and Young-Jai Park\footnote{E-mail address : yjpark@physics.sogang.
ac.kr}}}
\begin{center}
    {Department of Physics and Basic Science Research Institute}\\
   {Sogang University, C.P.O.Box 1142, Seoul 100-611, Korea}\\
\end{center}
\vspace{0.5cm}
\centerline{\Large Abstract}

  We investigate whether the gravitational thermodynamic properties of the 
scalar-tensor theory of gravity are affected by the conformal transformation 
or not. As an explicit example, we consider an electrically charged static 
spherical black hole in the 4-dimensional low energy effective theory of 
bosonic string.
        \newpage

   {\bf  I. Introduction}
                
   The conformal transformation of the action has been studied for various 
reasons. The usual actions in massless field theories are known to have 
conformal (Weyl) invariance, and ambiguities do not appear at the classical 
level [1]. Especially, in the scalar-tensor theories of gravity, one frequently
transforms  the action to an Einstein-Hilbert form for convenience of
calculation and so on. This is based on the fact that the conformal
transformation does not change the physics of the system, but just shrinks or
stretches the manifold [2].  However, we note that in many cases the surface 
terms in the gravitational action play an important role in the action 
principle, and these terms are affected by a conformal transformation. 

    The purpose of this paper is to study in detail the effect of a conformal 
transformation on the scalar-tensor theory of gravity.  In particular, we will
focus our attention on the total energy and the entropy of a gravitational 
system. These will be derived from both the action describing the system and 
the conformally transformed one. And we will compare them.  
In the next section, we carry out the formulation of the total energy of the
spatially bounded gravitational system. It is derived from both the 
nonminimally coupled action and the conformally transformed action (the 
minimal action) by using the gravitational Hamiltonians. In section III, the 
4-dimensional low energy effective theory of bosonic string is investigated as
an explicit example of the formulation. And the total energies observed in 
asymptotic region ($r \rightarrow \infty$) are calculated in both cases. In 
section IV, we compute the black hole entropy in the 4-dimensional low energy 
effective theory of bosonic string.  Section V contains the conclusion.
          \\        

              {\bf II. Formulation of the total energies}

  Consider a manifold  $(M,g_{ \mu \nu })$ with  the boundary $\partial M$
which consists  of initial  and  final  spacelike  surfaces
($\Sigma_{t^{\prime}}$ and  $\Sigma_{t^{\prime \prime}}$, respectively) and a
surface near infinity (timelike three-surface  $\Sigma^{\infty}$).
 We  assume that the spacelike
surfaces $\Sigma_{t}$ is orthogonal  to $\Sigma^{\infty}$ at the
two-dimensional  spacelike boundary $S^{\infty}_{t}$  of  $\Sigma_{t}$.
The  orthogonality means that  on the boundary $\Sigma^{\infty}$,
the timelike unit normal $n^{  \mu }$ to $\Sigma_{t}$ and the spacelike unit
  normal $r^{  \mu }$ to $\Sigma^{\infty}$ satisfy the relation
 $(n \cdot  r)|_{\Sigma^{\infty}} = 0$.
In other  words,  the product of  $S^{\infty}_{t}$  with segments of
timelike world lines orthogonal to $\Sigma_{t}$ at $S^{\infty}_{t}$
is the three-boundary $\Sigma^{\infty}$. 
           The  extrinsic curvatures of $\Sigma_{t}$  and
$\Sigma^{\infty}$ as embedded in  $M$ are  denoted $K_{ij}=h_i^k \nabla_k n_j$
and $\Theta_{ij}= \gamma_i^k \nabla_k  r_j$, respectively.  The extrinsic
curvature  of the two-dimensional  spacelike boundary  $S^{\infty}_{t}$
as  embedded  in $\Sigma_{t}$  is  $k_{ab}= \sigma_a^k  D_k r_b$, where $h_{ij}$, $\gamma_{ij}$ and $\sigma_{ab}$ are the induced metrics of $\Sigma_{t}$, 
$\Sigma^{\infty}$ and $S^{\infty}_{t}$, respectively.
The simbols $\nabla$ and $D$ denote covariant  differentiations with
respect to the metrics $g_{ \mu  \nu}$  and  $h_{ij}$, respectively.

   In the spacetime $M$, if a scalar field $\Phi$ is  nonminimally coupled to
scalar curvature, the action  describing the system can be written as
\begin{equation}
I = \int_{M} d^{4} x \sqrt{-g} \left(
               \frac{1}{16\pi}\Phi R + {\cal L}_{m}
                               \right)
        +\frac{1}{8\pi} \int_{\Sigma_{t}} d^{3} x \sqrt{h} \Phi K
   +\frac{1}{8\pi} \int_{\Sigma^{\infty}} d^{3} x \sqrt{-\gamma} \Phi \Theta,
\end{equation}
where ${\cal L}_{m}$ is the matter Lagrangian density which is assumed to
involve at most the first derivative and does not contain the gauge field.
The surface term is required so that the action yields the correct equations
of motion subject only to the condition that the boundary variables, which 
are induced three metrics and matter fields on the boundary, are held fixed.
Now we can obtain a new action of the Einstein-Hilbert form
through an appropriate conformal transformation from the action (1).
\begin{equation}
\tilde{I} = \int_{M} d^{4} x \sqrt{-\tilde{g}} \left(
            \frac{1}{16\pi} \tilde{R} + \tilde{{\cal L}}_{m}
                     \right)
  +\frac{1}{8\pi} \int_{\Sigma_{t}} d^{3} x \sqrt{\tilde{h}}\tilde{K}
  +\frac{1}{8\pi} \int_{\Sigma^{\infty}}
              d^{3} x \sqrt{-\tilde{\gamma}}\tilde{\Theta},
\end{equation}
where $g_{\mu \nu} =\Phi^{-1}(x) \tilde{g}_{\mu \nu}$.
       
 Recently, Hawking and Horowitz have proposed the general form of the total
energy for the spacetime with noncompact geometry as well as compact one [3].
They have shown that the boundary terms in the Hamiltonian come directly from
the boundary terms in the action rather than being put `by hand'.
If the static slices are labeled as $N_{0}=N$ on $\Sigma^{\infty}$,
the physical Hamiltonian derived from the action (2) is then
\begin{eqnarray}
 \tilde{H}_{p} & \equiv &  \tilde{H} - \tilde{H}_{0} \nonumber \\
       &=& \int_{\Sigma_{t}} (\tilde{N}\tilde{{\cal H}} + \tilde{N}^{i}\tilde{{\cal H}}_{i})
   - \int_{S^{\infty}_{t}}\left(\frac{1}{8 \pi}
\tilde{N} (\tilde{k} -\tilde{k}_{0})- 2\tilde{N}^{\mu}\tilde{p}_{\mu \nu}
 \tilde{r}^{\nu}/\sqrt{\tilde{h}} \right), \nonumber
\end{eqnarray}
where `0' denotes the reference background, and the following relation is 
used; 
\begin{eqnarray}
R = {}^{3}R + K_{\mu \nu}K^{\mu \nu} - K^{2}
           + 2\nabla_{\mu}(n^{\mu}\nabla_{\nu}n^{\nu})
          - 2\nabla_{\nu}(n^{\mu}\nabla_{\mu}n^{\nu}).
\end{eqnarray}
Given a solution, its total energy associated with the time translation 
$\tilde{t}^{\mu}$ is
\begin{eqnarray}
 \tilde{E} =   - \int_{S^{\infty}_{t}}d^{2}x \sqrt{\tilde{\sigma}}
                \left[\frac{1}{8 \pi}
       \tilde{N}\tilde{k}
       - 2\tilde{N}^{\mu}\tilde{p}_{\mu \nu}\tilde{r}^{\nu}/\sqrt{\tilde{h}}
                   \right]^{cl}_{0},
\end{eqnarray}
where $]^{cl}_{0}$ denotes evaluation at the classical solution minus evaluation for the
chosen reference background space.

  Now we generalize their procedure to the nonminimally coupled action (1).
In this case, the scalar field $\Phi$ is multiplied by curvature scalar and
 the total derivative terms in eq.(3) can not directly contribute to the
boundary terms.  But, after partial integrations, we obtain the appropriate
boundary terms as follows,
\begin{eqnarray}
 &~&  \frac{1}{16\pi} \int_{M} d^{4} x \sqrt{-g} \Phi R
              \nonumber  \\
  &=& \frac{1}{16\pi}\int_{M} d^{4} x  N\sqrt{h} [
            \Phi({}^{3}R + K_{\mu \nu}K^{\mu \nu} - K^{2})
   - 4n^{\mu}\partial_{\mu}\Phi K - 2D_{\mu}D^{\mu}\Phi]
                 \nonumber  \\
       &~&- \frac{1}{8\pi} \int_{\Sigma_{t}} d^{3} x \sqrt{h} \Phi K
   +\frac{1}{8\pi} \int_{\Sigma^{\infty}} d^{3} x \sqrt{-\gamma}
       [\Phi n^{\mu}n^{\nu}\nabla_{\nu}r_{\mu} + r^{\mu}\partial_{\mu}\Phi].
\end{eqnarray}
Substituting eq.(5) to the action (1), we can rewrite the action as following
\begin{eqnarray}
  I &=& \frac{1}{16 \pi} \int_{M} d^{4}x N\sqrt{h}
       [\Phi({}^{3}R + K_{\mu \nu}K^{\mu \nu} - K^{2})
   - 4n^{\mu}\partial_{\mu}\Phi K -2D_{\mu}D^{\mu}\Phi + 16 \pi {\cal L}_{m}]
            \nonumber  \\
 &~&+ \frac{1}{8 \pi}\int_{\Sigma^{\infty}} d^{3} N \sqrt{\sigma}
              (\Phi k + r^{\mu}\partial_{\mu}\Phi).
\end{eqnarray}
Note that a new boundary term $\int_{\Sigma^{\infty}}r^{\mu}\partial_{\mu}\Phi$
appears in eq.(6). As a result, the total energy derived from the action (1)
is
\begin{eqnarray}
 E  =  - \int_{S^{\infty}_{t}} d^{2}x \sqrt{\sigma}\left[\frac{1}{8 \pi}
            N(\Phi k + r^{\mu}\partial_{\mu}\Phi)
                  - 2N^{\mu}p_{\mu \nu}r^{\nu}/\sqrt{h}\right]^{cl}_{0}.
\end{eqnarray}
Comparing eq.(4) and eq.(7), one sees that the expressions of the total 
energy, which are computed from both the minimal and the nonminimal actions, 
are not equal to each other.  However, when $\Phi=1$, these become the same 
quantities. Note that the difference between $\tilde{E}$ and $E$ is mainly due
to the projection to the two-dimensional boundary of the gradient of the 
scalar field. In other words, the term $(\Phi k + r^{\mu}\partial_{\mu}\Phi)$ 
seems to play a role of the effective extrinsic curvature associated with the 
boundary $S^{\infty}_{t}$ in the nonminimal case.
             \\

 {\bf  III.  The 4-dimensional low energy effective theory of bosonic string}

   Now we shall apply the results in the previous section to the 
4-dimensional low energy effective theory of bosonic 
string as an explicit example.  The action of the theory is
\begin{eqnarray}
I &=& \int_{M} d^{4} x \sqrt{-g} e^{-2\phi} \left[
     \frac{1}{16 \pi}R +\frac{1}{4 \pi} \partial_{\mu}\phi \partial^{\mu}\phi
     -\frac{1}{4}F_{\mu \nu}F^{\mu \nu} \right] \nonumber  \\
    &&+\frac{1}{8\pi} \int_{\Sigma_{t}} d^{3} x \sqrt{h} e^{-2\phi}K
 +\frac{1}{8\pi} \int_{\Sigma^{\infty}} d^{3} x \sqrt{-\gamma}e^{-2\phi}\Theta,
\end{eqnarray}
where $\phi$ is the dilaton field, and $F$ is the field strength of the gauge
field $A$.  Through a conformal transformation 
$g_{\mu \nu} = e^{2\phi}\tilde{g}_{\mu \nu}$, we obtain a new action of the 
Einstein-Hilbert form as follows 
\begin{eqnarray}
\tilde{I} &=& \int_{M} d^{4} x \sqrt{-\tilde{g}} \left[
     \frac{\tilde{R}}{16 \pi} -\frac{1}{8 \pi}
                      \partial_{\mu}\phi \partial^{\mu}\phi
            -\frac{1}{4}e^{-2\phi}F_{\mu \nu}F^{\mu \nu} \right]
                           \nonumber \\
  && +\frac{1}{8\pi} \int_{\Sigma_{t}} d^{3} x \sqrt{\tilde{h}}\tilde{K}
  +\frac{1}{8\pi} \int_{\Sigma^{\infty}}
              d^{3} x \sqrt{-\tilde{\gamma}}\tilde{\Theta},
\end{eqnarray}
where the raising and lowering of indicies are carried out by the new metric 
$\tilde{g}_{\mu \nu}$. Varing the above two actions and solving the Einstein's
equations, we have obtained the electrically charged static spherical black 
hole solutions as follows
\begin{eqnarray}
ds^{2} = -(1-\frac{r_{H}}{r})(1+\frac{\alpha}{r})^{-2} dt^{2}
      +  (1-\frac{r_{H}}{r})^{-1} dr^{2} +r^{2} d\Omega_{2}^{2},
\end{eqnarray}
and
\begin{eqnarray}
d\tilde{s}^{2} = -(1-\frac{r_{H}}{r})(1+\frac{\alpha}{r})^{-1} dt^{2}
      +  (1-\frac{r_{H}}{r})^{-1} (1+\frac{\alpha}{r})dr^{2}
      + r^{2}(1+\frac{\alpha}{r}) d\Omega_{2}^{2},
\end{eqnarray}
respectively, where $\alpha \equiv Q^{2}/4\pi M$ and 
$r_{H} \equiv 2M(1- Q^{2}/8\pi M^{2})$ [4]. Here, $Q$ is the electric charge 
and $M$ is the ADM mass.  The corresponding dilaton and gauge fields are given
by
\begin{eqnarray}
e^{-2\phi} &=& (1 + \frac{\alpha}{r}), \nonumber  \\
A_{\mu} &=& \left(
           \frac{Q}{4\pi r}(1+\frac{\alpha}{r})^{-1} - \Phi_{H},~0,~0,~0
                                \right),
\end{eqnarray}
respectively, where $\Phi_{H} \equiv Q/8\pi M$ is the gauge fixing term chosen
on the condition of regularizing the gauge field at horizon $r = r_{H}$.

   Since both actions (8) and (9) contain the gauge field, we have another 
surface term $\int_{s^{\infty}_{t}} A_{t}p^{A}_{\mu} r^{\mu}$ in the process of
deriving the Hamiltonian, where $p_{A}^{\mu}$ is the momenta conjugate to the
gauge field $A_{\mu}$.   Thus the total energy for the minimal case (4) becomes
[5]
\begin{eqnarray}
 \tilde{E} = - \int_{s^{\infty}_{t}}d^{2}x \sqrt{\tilde{\sigma}}\left[
              \frac{1}{8 \pi}\tilde{N}\tilde{k}
       - 2\tilde{N}^{\mu}\tilde{p}_{\mu \nu}\tilde{r}^{\nu}/\sqrt{\tilde{h}}
         - A_{t}\tilde{p}^{A}_{\mu}\tilde{r}^{\mu}/\sqrt{\tilde{h}}
                   \right]^{cl}_{0}.
\end{eqnarray}
On the other hand, the total energy for the nonminimal case (7) becomes
\begin{eqnarray}
 E  =  - \int_{s^{\infty}_{t}} d^{2}x \sqrt{\sigma}\left[\frac{1}{8 \pi}
            N(\Phi k + r^{\mu}\partial_{\mu}\Phi)
                 - 2N^{\mu}p_{\mu \nu}r^{\nu}/\sqrt{h}
  - A_{t}p^{A}_{\mu} r^{\mu}/\sqrt{h}
                    \right]^{cl}_{0},
\end{eqnarray}
where  $\Phi \equiv e^{-2\phi}$. 

  Substituting eqs. (10 - 12) to (13) and (14), we obtain the total
energies observed in asymptotic region of the black hole as
\begin{eqnarray}
\tilde{E} = M - Q\Phi_{H} +\frac{\alpha}{2}~~{\rm and}
~~E = M -Q\Phi_{H} - \alpha,
\end{eqnarray}
respectively [6]. Here, we set $N=N_{0}=1,~\tilde{N}=\tilde{N}_{0}=1$. 
Note that this restriction gives the value of the Hamiltonian, which generates
unit time translations. 
And we have required that the reference background space is a static solution 
to the field equations ($p^{\mu \nu}|_{0},~p_{\Phi}|_{0},~p_{A}^{\mu}|_{0}$, 
and the constraints vanishes), and $\Phi |_{0} = 1$.

    In eq.(15), we can see that $\tilde{E}$ differs from the usual
total energy of the grand canonical ensemble, $M-Q\Phi_{H}$, by the facter
$\alpha/2$.  This is due to the dilaton charge $\cal{D}$ observed in infinite
region 
\begin{eqnarray}
  {\cal D} \equiv \frac{1}{ 4 \pi} \int_{\Sigma^{\infty}_{t}} d^{2}x
              \sqrt{\tilde{\sigma}} \tilde{r}^{\mu}\nabla_{\mu} \phi
              = \frac{\alpha}{2} + {\cal O}(r^{-1}).
\end{eqnarray}
Qualitatively, this result is originated as follows: although the dilaton 
charge is not an independent degree of freedom in the grand canonical ensemble 
($\alpha= Q^{2}/4\pi M$), this characterizes the black hole solution in the a
symptotic region [7]. So it seems natual for the dilaton charge to be
included in the total energy observed in asymptotic region.  On the other hand,
since the values of $\tilde{E}$ and $E$ differ by the facter $3\alpha/2$, we 
can say that the total energy observed in asymptotic region is changed by 
conformal transformation. 
          \\           
               
       {\bf V. The black hole entropy}

    In this section, we shall show that the black hole entropies computed from
two actions (8) and (9) are equal to each other at the semiclasical limit [8].
Calculating the entropy from the Einstein-Hilbert action (9) and
the corresponding black hole solution (11), we will show that the entropies 
calculated via well-known three methods are equal to each other.
It is important to note that we
use the results in eq.(15) for the entropies of the black holes with the 
Gibbons-Hawking method [8].

   Firstly, from the Bekenstein-Hawking relation [9,10], we obtain
\begin{eqnarray}
\tilde{S}_{m} = \frac{1}{4} A^{H}
          = 4 \pi M^{2}\left( 1 - \frac{Q^{2}}{8 \pi M^{2}} \right).
\end{eqnarray}
Secondly, from the first law of black hole thermodynamics, 
$T_{H}dS= dM -\Phi_{H} dQ$ [11] (the electrically charged black hole in 
eq.(11) is characterized by ($M,~Q,~J=0$)), the black hole entropy is given by
\begin{eqnarray}
\tilde{S}_{m} = \int \frac{dM}{\tilde{T}_{H}}
                     -\int \frac{\Phi_{H}}{\tilde{T}_{H}}dQ
            = 4 \pi M^{2} \left(1 - \frac{Q^{2}}{8 \pi M^{2}} \right),
\end{eqnarray}
where the Hawking temperature $\tilde{T}_{H}$ is $\tilde{T}_{H} = T_{H} = 
{\cal K}/2\pi =1/8\pi M$ and $\cal{K}$ is the surface gravity of the black 
hole.

   Next, let us compute the entropy with the method suggested by Gibbons and
Hawking [8].  The Euclidean action is given by
\begin{eqnarray}
\tilde{I}_{E}[\tilde{g},~A,~\phi] &=& -\left[\int_{Y} d^{4} x \sqrt{-\tilde{g}}
        \left(\frac{\tilde{R}}{16 \pi}
        -\frac{1}{8 \pi}\partial_{\mu}\phi \partial^{\mu}\phi \right)
 +\frac{1}{8\pi} \int_{\partial Y} d^{3} x \sqrt{\tilde{\gamma}}
                 (\tilde{\Theta}-\tilde{\Theta}_{0})\right]
                 \nonumber  \\
 &&+\frac{1}{4}\int_{Y} d^{4} x \sqrt{\tilde{g}}e^{-2\phi}F_{\mu \nu}F^{\mu \nu},
\end{eqnarray}
where the 4-dimensional Euclidean manifold $Y= \Re \times S^{1} \times S^{2}$,
and its boundary $\partial Y =S^{1}\times S^{2}$.  Then, the gravitational
partition function and the entropy are given by
\begin{eqnarray}
\tilde{Z}=\int [d\tilde{g}_{\mu \nu}][d A_{\mu}][d \phi] 
          e^{-\tilde{I}[\tilde{g},~A,~\phi]}
 \simeq  e^{-\tilde{I}[\tilde{g}_{c},~A_{c},~\phi_{c}]}, \nonumber  \\
\tilde{S}_{m} \simeq -\tilde{I}[\tilde{g}_{c},~A_{c},~\phi_{c}]
     + \frac{1}{\tilde{T}_{H}}\left( M - Q\Phi_{H} + \frac{\alpha}{2} \right),
\end{eqnarray}
where the subscript `c' denotes the saddle point of the Euclidean action and 
\begin{eqnarray}
\tilde{I}_{E} [\tilde{g}_{c},~A_{c},~\phi_{c}]
         = \frac{1}{2\tilde{T}_{H}}( M - Q\Phi_{H} + \alpha) +{\cal O}(r^{-1}).
\end{eqnarray}
Therefore, the black hole entropy is obtained as follows
\begin{eqnarray}
\tilde{S}_{m} \simeq  4 \pi M^{2} \left(1 - \frac{Q^{2}}{8 \pi M^{2}} \right).
\end{eqnarray}
This is the same as the results in eqs.(17) and (18) as expected.
Here, it must be emphasized that we did not use $(M - Q\Phi_{H})$ as the
total energy of the system, but $(M - Q\Phi_{H} +\alpha/2)$ derived in the 
previous section. Note that if we substitute $(M - Q\Phi_{H})$ in eq.(20), we 
could not have obtained the same black hole entropy with which evaluated in 
alternative ways.

     On the other hand, for the nonminimal case we can also compute the black 
hole entropy via Gibbons-Hawking method. The Euclidean action is  given by
\begin{eqnarray}
I_{E}[g,~A,~\phi] &=& -\left[\int_{Y} d^{4} x \sqrt{g}
        e^{-2\phi}\left(\frac{R}{16 \pi}
        +\frac{1}{4 \pi}\partial_{\mu}\phi \partial^{\mu}\phi \right)
 +\frac{1}{8\pi} \int_{\partial Y} d^{3} x \sqrt{\gamma}e^{-2\phi}
                 (\Theta-\Theta_{0})\right]
                 \nonumber  \\
 &&+\frac{1}{4}\int_{Y} d^{4} x \sqrt{g}e^{-2\phi}F_{\mu \nu}F^{\mu \nu}.
\end{eqnarray}
In the saddle point approximation, the first integration in eq.(23) does not
vanish via the equations of motion. This term gives another surface
term as follows
\begin{eqnarray}
-\frac{3}{16\pi} \int_{\partial Y} d^{3}x \sqrt{\gamma}(g_{rr})^{-1/2}
        \partial_{r}\left(1+ \frac{\alpha}{r} \right)
        = \frac{3\alpha}{4}\left(\frac{2\pi}{\cal{K}}\right) +{\cal O}(r^{-1}).
             \nonumber 
\end{eqnarray}
Then, we obtain the value of the action in saddle point approximation
\begin{eqnarray}
I_{E}[g_c,~A_c,~\phi_c] 
               =\frac{1}{2T_{H}}(M - Q\Phi_{H}-2\alpha) + {\cal O}(r^{-1}).
        \nonumber
\end{eqnarray}
As a result, the black hole entropy becomes
\begin{eqnarray}
S_{n} \simeq -I[g_c, A_c, \phi_c] + \frac{1}{T_{H}}(M-Q\Phi_{H}-\alpha)
                       \simeq 4\pi M^2 \left( 1- \frac{Q^2}{8\pi M^2} \right).
\end{eqnarray}
Therefore, we have shown that $S_{n} =\tilde{S}_{m}$.  From this result, one
recognize that the black hole entropy is not scaled by the conformal 
transformation up to the semiclassical limit.
                  \\

             {\bf  VI.  Conclusion}

    In this paper, we have explicitly examined whether the total energy and the
entropy of a gravitational system, in particular a black hole, are affected by
the conformal transformation or not.  The total energy is computed via 
deriving the gravitational Hamiltonian from the action. As a result, we have
shown that the total energy computed from the conformally transformed action 
(minimal ation) is different with that computed from the original action 
(nonminimal action) such that the extrinsic curvature associated with the 
two-dimentional boundary $S^{\infty}_{t}$, $\tilde{k}$ in the former is 
replaced with the {\it effective
extrinsic curvature} $(\Phi k + r^{\mu}\partial_{\mu}\Phi)$ in the latter. 

    In the 4-dimensional low energy effective theory of bosonic string, the
total energies observed in asymptotic region of the electrically charged 
static spherical black hole are given in eq.(15). They differ by the facter 
$3/2 \alpha$, and the facter is related with the dilaton
charge ${\cal D}$. So, we can say that the total energy observed in asymptotic
region is scaled by the conformal transformation. Also, we  note that 
the total energies are different from the usual total energy of the grand 
canoniocal ensemble, $(M-Q\Phi_{H})$. In order to understand it we have
to consider two factors related each other. One is the method computing
total energy. We have considered a timelike boundary, and then it is sent
back to infinite region. Another is that although the dilaton charge is 
not the independent degree of freedom in the grand canonical ensemble, the 
black hole solution is characterized by the dilaton charge as well as the
ADM mass and the electric charge in asymptotic region. In other words, owing 
to the method, the effect of the dilaton charge is contained in the total 
energy of the gravitational system.

    In the case of the action of the Einstein-Hilbert form, the black hole
entropy is calculated in the well-known three methods. And all of these three
methods have yielded the same expressions for entropy.  On the process of the
calculation via the Gibbons-Hawking method, it is the important point that
we did not use $(M-Q\Phi_{H})$ as the total energy of the black hole, but
$(M-Q\Phi_{H} + \alpha/2)$. On the other hand, when the black hole entropy is
computed from the nonminimally coupled action via the Gibbons-Hawking method, 
also we have used the result in eq.(15) as the total energy.  As a result, we 
have shown that this gives the same expression for the entropy with that 
calculated from the Einstein-Hilbert form. Furthermore, we have obtained the 
same Hawking temperature in two cases. Of couse, the local (Tolman redshift) 
temperatures, $T(r) =(-g_{tt})^{-1/2} T_{H}$ and $\tilde{T}(r) =
(-\tilde{g}_{tt})^{-1/2}\tilde{T}_{H}$, are different each other.  

   In conclution, the conformal transformation  changes local geometry, while 
Hawking temperature $T_{H}$ and black hole entropy $S$ are global quantities. 
Therefore, the values of $T_{H}$ and $S$ are invariant under the conformal 
transformation [12, 13].
                   \\

      {\large \bf \rm Acknowledgments}

   This work was supported by Basic Science Research Institute Program,
Ministry of Education, Project No. 95-2414.
                     \\

      {\large \bf \rm References}

\begin{description}
\item{[1]} S. B. Treiman, R. Jackiw, B. Zumino and E. Witten,
           {\it Current Algebra and Anomalies} (Princeton University Press,
           Princeton, 1985).
\item{[2]} N. D. Birrell and P. C. W. Davies, 
	   {\it Quantum Fields in Curved Space} (Cambridge University Press,
	   Cambridge, 1982).
\item{[3]} S. W. Hawking and G. T. Horowitz, ``The Gravitational Hamiltonian,
           Action, Entropy and Surface terms", gr-qc/9501014.
           J. D. Brown and J. W. York, Jr., Phys. Rev. D{\bf 47}, 1407(1993).
\item{[4]} G. W. Gibbons and K. Maeda, Nucl. Phys. B{\bf 298}, 741(1988).
           I. Ichinose and H. Yamazaki, Mod. Phys. Lett. A{\bf 4}, 1509(1989). 
\item{[5]} S. W. Hawking and S. F. Ross, ``Duality between Electric
            and Magnetic Black Holes", hep-th/9504019.
\item{[6]} Compare with the result of J. Katz, D. Lynden-Bell, and W. Israel, 
           Class. Quantum Grav. {\bf 5}, 971(1988).
\item{[7]} D. Garfinkle, G. Horowitz and A. Strominger, Phys. Rev. D{\bf 43},
           3140(1991).
\item{[8]} G. W. Gibbons and S. W. Hawking, Phys. Rev. D{\bf 15}, 2752(1977).
           For rotating black holes, J. D. Brown, E. A. Martinez and
           J. W. York, Jr., Phys. Rev. Lett. {\bf 66}, 2281(1991).
\item{[9]} J. D. Bekenstein, Phys. Rev. D{\bf 7}, 2333(1973); {\it ibid}
           D{\bf 9}, 3292(1974).
\item{[10]} S. W. Hawking, Commun. Math. Phys. {\bf 43}, 199(1975).
\item{[11]} J. W. Bardeen, B. Carter and S. W. Hawking, Comm. Math. Phys.
            {\bf 31}, 161(1973).
\item{[12]} R. Kallosh, T. Ort$\acute{i}$n and A. Peet, Phys. Rev. D{\bf 47},
            5400(1993).
\item{[13]} K. C. K. Chan, J. D. E. Creighton and R. B. Mann, Phys. Rev. 
	    D{\bf 54}, 3892(1996).
\end{description}
\end{document}